\documentclass{ar2e}
\input epsfig.sty

\jname{Annu. Rev. Astron. Astrophys.}
\jyear{2002}
\jvol{40}
\ARinfo{Preprint astro-ph/0111170}

\def\simg{\mathrel{\hbox{\rlap{\lower.55ex \hbox {$\sim$}}
	   \kern-.3em \raise.4ex \hbox{$>$}}}}
\def\siml{\mathrel{\hbox{\rlap{\lower.55ex \hbox {$\sim$}}
	   \kern-.3em \raise.4ex \hbox{$<$}}}}
\def\Mesz{M\'esz\'aros~}
\def\Pacz{Paczy\'nski~}
\def\nsns{NS-NS~}
\def\bhns{BH-NS~}

\def\beq{\begin{equation}}
\def\enq{\end{equation}}
\def\bea{\begin{eqnarray}}
\def\ena{\end{eqnarray}}
\def\bec{\begin{center}}
\def\enc{\end{center}}
\def\etal{{et al.~}}
\def\msun{M_\odot}
\def\eps{\epsilon}
\def\varep{\varepsilon}
\def\Omj{\Omega_j}
\def\Mbh{M_{bh}}

\newcommand{\boxsize}{0.89\textwidth}

\begin{document}


\title{Theories of Gamma-Ray Bursts}

\markboth{P. \Mesz }{Theories of Gamma-Ray Bursts }

\author{P. \Mesz
\affiliation{
Dept. of Astronomy \& Astrophysics, 525 Davey Laboratory, \\
Pennsylvania State University, University Park, PA 16803}}

\maketitle

\section*{Abstract}
The gamma ray burst phenomenon is reviewed from a theoretical point of view, 
with emphasis on the fireball shock scenario of the prompt emission and the 
longer wavelenght afterglow. Recent progress and issues are discussed, including 
spectral-temporal evolution, localizations, jets, spectral lines, environmental 
and cosmological aspects, as well as some prospects for future experiments in
both electromagnetic and non-electromagnetic channels.

\noindent
{\it Keywords:} {Gamma-ray bursts - gamma-rays - high energy - cosmology - neutrinos}

\section{Introduction}
\label{sec:intro}

Gamma-Ray Bursts (GRB) were first detected in the late 1960's by military satellites
monitoring for compliance with the nuclear test ban treaty. This became public
information only several years later, with the publication of the results from the 
Vela satellites \cite{klebesadel73}, which were quickly confirmed by data from the 
Soviet Konus satellites \cite{mazets74}. Their nature and origin remained thereafter 
a mystery for more than two decades, largely due to the fact that during this period
they remained detectable only for tens of seconds, almost exclusively at gamma-ray 
energies (e.g. \citen{hurley91}), with occasional reports at X-ray energies 
(e.g. \citen{murakami88,yoshida89,connors98}). Various satellites continued to 
accumulate data on hundreds of GRB over the years, attracting an increasing amount
of attention and leading to a large variety of theoretical models (e.g. 
\citen{ruderman75,liang89}).

\begin{figure}[ht]
\begin{center}
\begin{minipage}[t]{0.5\textwidth}
\epsfxsize=\boxsize
\epsfbox{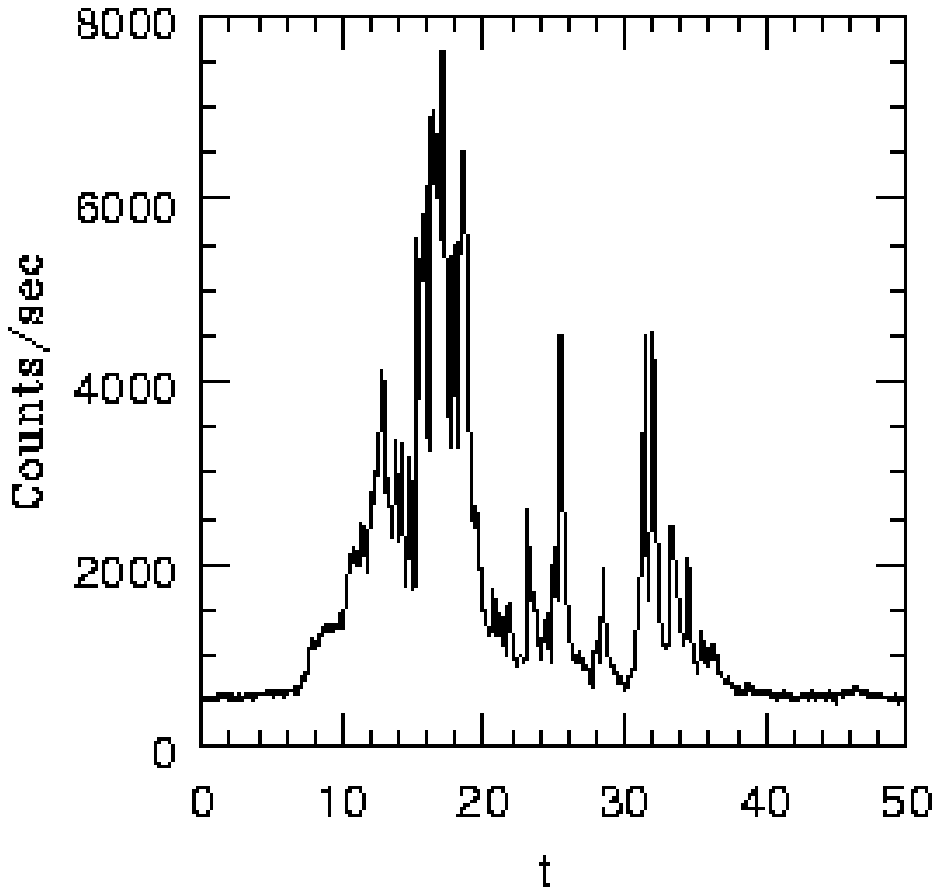}
\end{minipage}
\hspace{5mm}
\begin{minipage}[t]{0.45\textwidth}
\vspace*{-6cm}
\caption{Typical GRB lightcurve observed with BATSE, showing photon count rate 
(0.05-0.5 MeV) versus time (s). No $\gamma$-rays are detected either before or 
after the burst trigger \cite{fm95}. }
\label{fig:batselc}
\end{minipage}
\end{center}
\end{figure}

A new era in GRB research opened in 1991 with the launch of the Compton 
Gamma-Ray Observatory (CGRO), whose ground-breaking results have been summarized 
in \citen{fm95}.  The most significant results came from the all-sky survey by 
the Burst and Transient Experiment (BATSE) on CGRO, which recorded over 2700
bursts, complemented by data from the OSSE, Comptel and EGRET experiments. BATSE's
earliest and most dramatic result was that it showed that GRB were essentially 
isotropically distributed in the sky, with no significant dipole or quadrupole 
moments, suggesting a cosmological distribution \cite{meeg92}. The spectra were 
non-thermal, the number of photons per unit photon energy varying typically as 
a power-law $N(\epsilon)\propto \epsilon^{-\alpha}$, where $\alpha \sim 1$ at low 
energies changes to $\alpha\sim 2-3$ above a photon energy $\epsilon_{0}\sim 0.1-1$ 
MeV \cite{band93}. This spectral power law dependence was found to extend in 
several bursts up to at least GeV energies \cite{schneid95,hur94}.  The gamma-ray 
light curves show a time dependence ranging from a smooth, fast-rise and 
quasi-exponential decay, through curves with several peaks, to variable curves 
with  many peaks, and substructure sometimes down to milliseconds 
(Fig. \ref{fig:batselc}). 
The durations at MeV energies range from $10^{-3}$ s to about $10^3$ s , 
with a well-defined bimodal distribution for bursts longer or 
shorter than $t_b \sim 2$ s \cite{kou93}. There is also an anti-correlation between 
spectral hardness and duration, the short one being harder, e.g. \cite{fm95}.
The pulse distribution is complex, and the time histories of the emission as a 
function of energy can provide clues for the geometry or physics of the emitting 
regions (e.g. \citen{fen98,belo98}). The results from BATSE sharpened 
the debate on whether the GRB were of a galactic or extragalactic origin, e.g. 
\cite{lamb95,pacz95}, but the accumulating evidence increasingly swung the
balance in favor of the cosmological interpretation.

A decisive watershed was reached in 1997, when the Italian-Dutch satellite 
Beppo-SAX succeeded in obtaining the first high resolution x-ray images 
\cite{cos97} of the fading afterglow of a burst, GRB 970228, which had been
expected on theoretical grounds. This discovery was promptly followed by an 
increasing list of other burst detections by Beppo-SAX, at the approximate 
rate of 10 per year. 
These X-ray detections, after a 4-6 hour delay needed for data processing, 
led to arc-minute accuracy positions, which finally made possible the optical
detection and the follow-up of the GRB afterglows at longer wavelengths (e.g.
\citen{jvp97,fra97}). This paved the way for the measurement of redshift distances, 
the identification of candidate host galaxies, and the confirmation that they were 
at cosmological distances (\citen{metz97,kul99b}, etc.). Over 40 GRB afterglows 
have been located as of late 2001 in X-rays and optical, and more than a dozen in
radio \cite{fra99,weiler02}. Some afterglows have been followed over time scales of 
many months to over a year, and in the majority of cases (over 30) they have also
resulted in the identification of the likely host galaxy 
\cite{bloom01,djorgovski01}. A recent review of the observations and phenomenology 
of GRB afterglows is in \citen{jvp00}.

\section{Gamma-ray burst phenomenology: the fireball shock scenario}
\label{sec:gamma}

At cosmological distances the observed GRB fluxes imply energies of 
$\siml 10^{54}$ erg, if the emission is isotropic (see however \S \ref{sec:jet}), 
and from causality this must be liberated inside regions whose size is
$\siml 100$ kilometers on time scales $\siml$ seconds. Independently of 
the nature and details of the progenitor and the trigger, such an intense, 
localized and brief explosion implies the formation of an $e^\pm,\gamma$ fireball 
\cite{cavrees78}. In the context of a cosmological model, the fireball would
be expected to expand relativistically \cite{pac86,goo86,pac90}. This hypothesis
is natural, since most of the spectral energy is observed at $\simg 0.5$ MeV, 
so the optical depth against $\gamma\gamma \to e^\pm$ is huge, and the expansion
follows from the highly super-Eddington value of the luminosity. Since many bursts 
emit a large fraction of their luminosity at photon energies $\eps_\gamma \gg 1$ 
MeV, the flow must somehow be able to avoid the process $\gamma\gamma \to e^\pm$
degrading the observed photons to just below 0.511 MeV. A highly relativistic
expansion is, in fact, strongly supported by the fact that it provides a natural 
explanation for the observed photons with $\eps_\gamma \gg $ 0.5 MeV 
\cite{fen93-2gam,hb94}. This is because in this case the relative angle at which 
the photons collide must be less than the inverse of the bulk Lorentz factor 
$\gamma^{-1}$ and the effective threshold energy for pair production is 
correspondingly reduced.  Roughly, the Lorentz factor must satisfy
\beq
\gamma \simg 10^2 (\eps_{\gamma}/{\rm 10 GeV})^{1/2}(\eps_t /{ \rm MeV})^{1/2}~,
\enq
in order for photons with $\eps_\gamma \simg 10$ GeV to escape annihilation
against target photons with $\eps_t \sim 1$ MeV. (A more detailed calculation 
is in \citen{lithwick01}).

From general considerations \cite{sp90}, a relativistic outflow arising
from an initial energy $E_o$ imparted to a mass $M_o << E_0/c^2$ starting out from
a radius $r_l$ leads to an expansion, as the gas converts its internal energy
into bulk kinetic energy.  Initially the bulk Lorentz $\gamma \simeq r/r_l 
\propto r$, while the comoving temperature drops $\propto r^{-1}$. Clearly,
$\gamma$ cannot increase beyond $\gamma_{max} \sim \eta \sim E_o/M_o c^2$,
which occurs at a saturation $r_s \simg r_l \eta$, beyond which the flow
continues to coast with $\gamma \sim \eta \sim $ constant.
The simplicity of the original fireball picture, however, led to some serious 
difficulties. Among these are that the expansion of the fireball should lead to
a conversion of most of its internal energy into kinetic energy of the
entrained baryons, rather than into photon luminosity, hence it would be 
energetically very inefficient. Furthermore, it would produce a quasi-thermal 
photon spectrum, instead of the observed power-law spectra; and the typical time 
scales over which these photons escape is comparable to that during which the 
flow makes a transition to optical thinness (milliseconds), which could not 
explain the many events lasting much longer than that. 

This efficiency, timescale and spectrum problems can be solved with the 
{\it fireball shock} model, in its external \cite{rm92} and internal \cite{rm94} 
versions. This is based on the fact that shocks are likely to occur in such an 
outflow, and if these occur after the fireball has become optically thin, these 
shocks would reconvert the kinetic energy of the baryons into nonthermal particle 
and photon energy. 

{\it External} shocks \cite{mr93a} will occur, unavoidably, in any outflow of 
total energy $E_o$ in an external medium of average particle density $n_o$ at a 
radius and on a timescale 
\bea
r_{dec} \sim & 10^{17} E_{53}^{1/3} n_o^{-1/3} \eta_2^{-2/3} 
         ~{\rm cm}~, \cr
t_{dec} \sim & r_{dec}/(c\gamma^2) \sim 3\times 10^2  E_{53}^{1/3} n_o^{-1/3}
                                                     \eta_2^{-8/3} ~{\rm s }~,
\label{eq:rdec}
\ena
where (in the impulsive, or thin shell approximation) the lab-frame energy of 
the swept-up external matter ($\gamma^2 m_p c^2$ per proton) equals the initial 
energy $E_o$ of the fireball, and $\eta=\gamma = 10^2\eta_2$ is the final bulk 
Lorentz factor of the ejecta. 

The external shock synchrotron spectra \cite{mr93a,ka94b} and combined 
synchrotron-IC spectra \cite{mlr93,mrp94} reproduce in a general manner the 
observed gamma-ray spectral properties, as do the predicted spectral-temporal 
correlations (\citen{sanapi96,pm98b,dermer99,bd00}; c.f. \citen{liang99}). 
(However, internal shocks present an alternative for the brief burst of
gamma-ray emission, motivated by variability issues, see below). External shocks 
also serve as the model of choice for the afterglow radiation (\S \ref{sec:after}).
The typical observer-frame dynamic time of the shock is 
$t_{dec} \sim r_{dec}/c \gamma^2 \sim$ seconds, for typical parameters, 
and $t_b \sim t_{dec}$ would be the burst duration (the
impulsive assumption requires that the initial energy input occur in a time
shorter than $t_{dyn}$). Variability on timescales shorter than $t_{dec}$
may occur on the cooling timescale or on the dynamic timescale for 
inhomogeneities in the external medium, but this is not widely favored
for reproducing highly variable profiles. (\citen{sapi98}; c.f. \citen{dm99}).
They could, however, reproduce bursts with several peaks \cite{pm98a} and may 
therefore be applicable to the class of long, smooth bursts.

{\it Internal} shocks \cite{rm94} address another problem, posed by some
of the rapidly variable $\gamma$-ray light curves, which for total durations 
of tens to hundreds of seconds are, sometimes, endowed with variability down 
to milliseconds or less \cite{fm95}. One ingredient in solving this problem is 
to postulate a ``central engine" \cite{fen93} which ejects energy at a variable 
rate. This could be, e.g. magnetic flares in a transient accretion disk around a 
central compact object resulting from the disruption of a merging compact binary 
\cite{napapi92}. By itself, such a variable central engine is however not 
enough to explain the variable light curves, since a relativistic outflow is 
inevitable, and even if intermittent, this outflow will be on average optically  
thick to Compton scattering out to very large radii, leading to a smoothing-out 
of the light curve. This difficulty, however, is solved with the introduction of
the internal shock model \cite{rm94}, in which the time-varying outflow from the 
central engine leads to successive shells ejected with different Lorentz factors. 
Multiple shocks form as faster shells overtake slower ones, and the crucial point 
is that for a range of plausible parameters, this occurs above the Compton 
photosphere.  These shocks are called internal because they arise from the flow 
interacting with itself, rather than with the external environment. 

One can model the central engine outflow as a wind of duration $t_w$, 
whose average dynamics is similar to that of the impulsive outflows described 
previously, with an average lab-frame luminosity $L_o= E_o/t_w$ and average mass 
outflow $\dot M_o$, and mean saturation Lorentz factor $\gamma \sim \eta=
L_o/ {\dot M_o c^2}$. Significant variations of order $\Delta\gamma \sim \gamma 
\sim \eta$ occurring over timescales $t_{var}\ll t_w$ will lead then to internal 
shocks \cite{rm94} at radii $r_{dis}$ above the photosphere $r_{phot}$,
\bea
r_{dis} \sim & c t_{var} \eta^2 \sim 3\times 10^{14} t_{var} \eta_2^2 ~
                                                           {\rm cm}, \cr
r_{phot} \sim & {\dot M}\sigma_T /(4\pi m_p c\eta^2)\sim 10^{11} L_{50}\eta_2^{-3}~
                                                           {\rm cm}. 
\label{eq:rdis}
\ena
The above assumes the photosphere to be above the saturation radius 
$r_s\simeq r_o \eta$, so that most of the energy comes out in the shocks, rather
than in the photospheric quasi-thermal component (such photospheric effects 
are discussed in \citen{mr00b}). For shocks above the photosphere, large
observable $\gamma$-ray variations are possible on timescales $t_{var} \simg
t_{var,min}\sim 10^{-3}(M_{c}/\msun)^{3/2}$, for an outflow originating from a 
central object of mass $M_c$ at radii $\simg r_o\sim c t_{var,min}/2 \pi$.
The internal shock  model was specifically designed to allow an arbitrarily 
complicated light curve \cite{rm94} on timescales down to ms, the optically 
thin shocks producing the required non-thermal spectrum. Numerical calculations 
\cite{kobayashi99,daigne00,spada00} confirm that the light curves can indeed be
as complicated as observed by BATSE in extreme cases. (By contrast, in external 
shocks the variations are expected to be smoothed out by relativistic time delays,
e.g. \citen{sapi98}, at most a few peaks being possible, e.g. 
\citen{pm98a}. An alternative view invoking large variability from blobs in 
external shocks is discussed by \citen{dm99}). The observed power density spectra 
of GRB light curves \cite{belo00} provide an additional constraint on the dynamics 
of the shell ejection by the central engine and the efficiency of internal 
shocks \cite{spada00,guetta01}.

When internal shocks occur, these are generally expected to be followed 
\cite{mr94,mr97a} by an external shock, a sequential combination sometimes 
referred to as the internal-external shock scenario \cite{piran98}. 
The GRB external shocks, similarly to what is observed in supernova remnants, 
consist of a forward shock or blast wave moving into the external medium ahead 
of the ejecta, and a reverse shock moving back into the ejecta as the latter 
is decelerated by the inertia  the external medium.  The internal shocks would 
consist of forward and reverse shocks of a more symmetrical nature. As in 
interplanetary shocks studied with spacecraft probes, the internal and external 
shocks in GRB are tenuous, and expected to be collisionless, i.e. mediated by 
chaotic electric and magnetic fields.  The minimum random Lorentz factor of 
protons going through 
the shocks should be comparable to the relative bulk Lorentz factor, while that 
of the electrons may exceed this by a factor of up to the ratio of the proton to 
the electron mass. The energy of the particles can be further boosted by diffusive 
shock acceleration \cite{be87} as they scatter repeatedly across the shock 
interface, acquiring a power law distribution $N(\gamma_e)\propto \gamma_e^{-p}$, 
where $p\sim 2-3$. In the presence of turbulent magnetic fields  built up behind 
the shocks, the electrons produce a synchrotron power-law radiation spectrum 
\cite{mr93a,rm94} similar to that observed \cite{band93}, while the inverse 
Compton (IC) scattering of these synchrotron photons extends the spectrum into the 
GeV range \cite{mrp94}. Comparisons of a synchrotron hypothesis for the MeV 
radiation with data have been made by, e.g. 
\citen{tav96,preece00,eiclev00,mr00b,medvedev00,panmesz00,lloyd01a}.
The effects of pair production and inverse Compton on the prompt spectra
are discussed in \S \ref{sec:pairs}.

It is worth stressing that the fireball shock scenario, whether internal or 
external, is fairly generic: it is largely independent of the details of the 
progenitor. Although it is somewhat geometry dependent, the central engine 
generally lies enshrouded and out of view inside the optically thick outflow. 
Even after the latter becomes optically thin, the progenitor's remnant emission 
should be practically undetectable, compared to the emission of the fireball 
shock which is its main manifestation (see, however, \S \ref{sec:cosm}).

\section{Blast Wave Model of GRB afterglows}
\label{sec:after}

The external shock becomes important when the inertia of the swept up external 
matter leads to an appreciable slowing down of the ejecta. As the fireball 
continues to plow ahead, it sweeps up an increasing amount of external matter, 
made up of interstellar gas plus possibly gas which was previously ejected by the 
progenitor star. As the external shock builds up, for high radiative efficiency
its bolometric luminosity rises approximately as $L\propto t^2$. This follows 
from equating in the contact discontinuity frame the kinetic flux $L/4\pi r^2 $ 
to the external ram pressure $\rho_{ext} \gamma^2$ during the initial phase 
while $\gamma\sim$ constant, $r\propto t$ (\citen{rm92}; see also \citen{sari98}).
After peaking, or plateauing in the thick shell limit, as the Lorentz factor 
decreases one expects a gradual 
dimming $L\propto t^{-1+q}$ (from energy conservation $L\propto E/t$ under 
adiabatic conditions, $q$ takes into account radiative effects or bolometric 
corrections). At the deceleration radius (\ref{eq:rdec}) the fireball energy 
and the bulk Lorentz factor decrease by a factor $\sim 2$ over a timescale 
$t_{dec}\sim r_{dec}/(c\gamma^2)$, and thereafter the bulk Lorentz factor 
decreases as a power law in radius,
\beq
\gamma \propto r^{-g}\propto t^{-g/(1+2g)}~,~r\propto t^{1/(1+2g)},
\label{eq:gamma}
\enq
with $g=(3,3/2)$ for the radiative (adiabatic) regime, in which
$\rho r^3 \gamma \sim$ constant ($\rho r^3 \gamma^2 \sim$ constant).
At late times, a similarity solution \cite{bm76a,bm76b} solution with $g=7/2$ 
may be reached.  
The spectrum of radiation is likely to be due to synchrotron radiation, whose 
peak frequency in the observer frame is $\nu_m \propto \gamma B' \gamma_e^2$,
and both the comoving field $B'$ and the minimum electron Lorentz factor 
$\gamma_{e,min}$ are likely to be proportional to $\gamma$ \cite{mr93a}. This 
implies that as $\gamma$ decreases, so will $\nu_m$, and the radiation will move 
to longer wavelengths. Consequences of this are the expectation that the burst 
would leave a radio remnant \cite{pacro93} after some weeks, and before that an 
optical \cite{ka94b} transient.

The first self-consistent afterglow calculations \cite{mr97a} took into account 
both the dynamical evolution and its interplay with the relativistic particle
acceleration and a specific relativistically beamed radiation mechanism 
resulted in quantitative predictions for the entire spectral evolution,
going through the X-ray, optical and radio range. For a spherical fireball 
advancing into an approximately smooth external environment, the bulk 
Lorentz factor decreases as in inverse power of the time (asymptotically 
$t^{-3/8}$ in the adiabatic limit), and the accelerated electron minimum random 
Lorentz factor and the turbulent magnetic field also decrease as inverse power 
laws in time. The synchrotron peak energy corresponding to the time-dependent 
minimum Lorentz factor and magnetic field then move to softer energies as 
$t^{-3/2}$. These can be generalized in a straightforward manner when in the
radiative regime, or in presence of density gradients, etc.. The radio  spectrum 
is initially expected to be self-absorbed, and becomes optically thin after a 
few days. For times beyond about one hour the dominant radiation  is from the 
forward shock, for which the flux at a given frequency and the synchrotron peak 
frequency decay as \cite{mr97a}
\beq
F_\nu \propto t^{-(3/2)\beta}~~,~~\nu_m\propto t^{-3/2}~,
\label{eq:Fnu}
\enq
as long as the expansion is relativistic. This is referred to as the ``standard" 
(adiabatic) model, where $g=3/2$ in equ. [\ref{eq:gamma}] and $\beta$ is the 
photon spectral energy slope ($F_\nu\propto \nu^{-\beta}$). The transition to the 
non-relativistic regime
has been discussed, e.g. by \citen{wrm97,dailu99,livwax00}.  More generally
\cite{mr99} the relativistic forward shock flux and frequency peak are given by
$F_\nu\propto  t^{[3-2g(1+2\beta)]/(1+2g)}$ and $\nu_m\propto t^{-4g/(1+2g)}$. 
(A reverse shock component is also expected \cite{mr93b,mr97a},  with high 
initial optical brightness but much faster decay rate than the forward shock, 
see \S \ref{sec:reverse}).
It is remarkable, however, that the simple ``standard" model 
where reverse shock effects are ignored is a good approximation for modeling
observations starting a few hours after the trigger, as during 1997-1998.

\begin{figure}[ht]
\begin{center}
\begin{minipage}[t]{0.5\textwidth}
\epsfxsize=\boxsize
\epsfbox{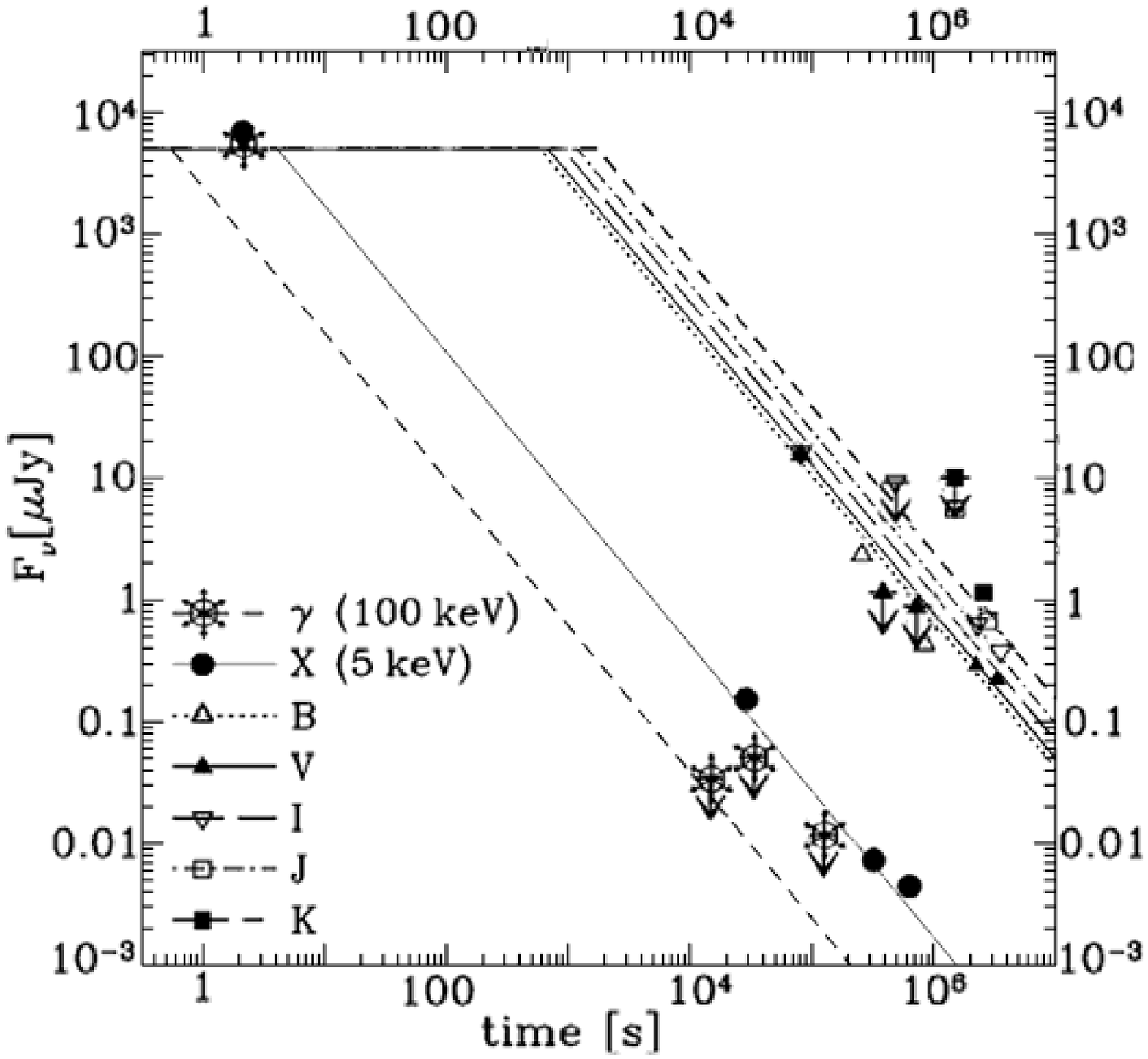}
\end{minipage}
\hspace{5mm}
\begin{minipage}[t]{0.45\textwidth}
\vspace*{-6cm}
\caption{Comparison \cite{wrm97} of the observed light curves of the afterglow 
of GRB 970228 at various wavelengths with the simple blast wave model predictions 
\cite{mr97a}. }
\label{fig:lc970228}
\end{minipage}
\end{center}
\end{figure}

The predictions of the fireball shock afterglow model \cite{mr97a} were made 
in advance of the first X-ray detections by Beppo-SAX \cite{cos97} allowing
subsequent follow-ups \cite{jvp97,metz97,fra99} over different wavelengths,
which showed a good agreement with the standard model, e.g. 
\cite{vie97a,wrm97,tav97,wax97a,rei97} 
(Fig. \ref{fig:lc970228}).  
The comparison of increasingly sophisticated versions of this theoretical model
(e.g. \citen{spn98,wiga99,piran99,dermer00,granot00a}) against an increasingly 
detailed array of observations (e.g. as summarized in \citen{jvp00}) has provided 
confirmation of this generic fireball shock model of GRB afterglows.  

A snapshot spectrum of the standard model at any given time consists of a 
three-segment power law with two breaks. At low frequencies there is a steeply 
rising synchrotron self-absorbed spectrum up to a self-absorption break $\nu_a$,
followed by a +1/3 energy index spectrum up to the synchrotron break $\nu_m$
corresponding to the minimum energy $\gamma_m$ of the power-law accelerated
electrons, and then a $-(p-1)/2$ energy spectrum above this break,
for electrons in the adiabatic regime (where $\gamma_e^{-p}$ is the electron
energy distribution above $\gamma_m$). A fourth segment is expected at energies 
above where the electron cooling time becomes short compared to the expansion 
time, with a spectral slope $-p/2$ above that, with a corresponding ``cooling" 
break $\nu_b$ \cite{mrw98,spn98}. The observations (e.g. \citen{jvp00}) are 
compatible with an electron spectral index $p\sim 2.2-2.5$ \cite{gallant99},
which is typical of shock acceleration, e.g. \citen{wax97a,spn98,wiga99}, etc.  
As the remnant expands the photon spectrum moves to lower frequencies, and the 
flux in a given band decays as a power law in time, whose index can change as 
breaks move through it.
Snapshot spectra have been deduced by extrapolating measurements at different 
wavelengths and times, and assuming spherical symmetry and using the model time 
dependences \cite{wax97b,wiga99}, fits were obtained for the different physical 
parameters of the burst and environment, e.g. the total energy $E$, the magnetic 
and electron-proton coupling parameters ${\eps}_B$ and ${\eps}_e$ and the external 
density $n_o$ (see right panel of Figure \ref{fig:jet0510}).
These lead to typical values $n_o\sim 10^{-2}-10$ cm$^{-3}$, $\eps_B\sim 10^{-2}$, 
$\eps_e\sim 0.1-0.5$ and $E\sim 10^{52}-10^{54}$ ergs (if spherical; but see
\S \ref{sec:jet}).

\section{Standard Model Developments and Issues}
\label{sec:postaft}

The standard afterglow model is based on the following approximations: 
a) spherical outflow;
b) a homogeneous external medium $n\sim n_o$;
c) highly relativistic expansion in the adiabatic approximation;
d) an impulsive energy input $E_o$ and a single $\gamma_o=\eta=E_o/M_o c^2$;
e) line of sight scaling relations assumed valid for the entire visible hemisphere;
f) time-independent shock acceleration  parameters $p$, $\varep_B$, $\varep_e$
(electron energy index, magnetic to proton  and electron to proton energy ratios);
g) only the forward shock radiation is included.
The significant success of this model in explaining many of the observations 
in the first years after GRB 970228 indicates that these approximations are 
robust, at least in a broad sense and over a range of timescales. However, they 
are clearly simplifications, and are expected
to be appropriate only within certain limits.

\subsection{Density, Angle and Time-dependent Injection}
\label{sec:refresh}
Departures from the simplest standard model occur, e.g. if the external medium 
is inhomogeneous. For instance, for $n \propto r^{-d}$, the energy 
conservation condition is $\gamma^2 r^{3-d} \sim$ constant, which changes 
significantly the temporal decay rates \cite{mrw98}.  Such a power law dependence 
is expected if the external medium is a wind, say from an evolved progenitor star,
and light curve to some bursts fit better with such a hypothesis \cite{chev00},
whereas in many objects a homogeneous medium seems a better fit 
\cite{fra01,pankum01b} (for a critical discussion see \citen{chevalier01}).
Another obvious non-standard effect is departures 
from a simple impulsive injection approximation (i.e. an injection which is not 
a delta or a top hat function with a single value for $E_o$ and $\gamma_o$ in 
time). An example is if the mass and energy injected during the burst duration 
$t_w$ (say tens of seconds) obeys $M(>\gamma) \propto \gamma^{-s}$, 
$E(>\gamma)\propto \gamma^{1-s}$, i.e. more energy emitted with lower Lorentz 
factors at later times, but still shorter than the gamma-ray pulse duration 
(refreshed shocks).
This would drastically change the temporal decay rate and extend the afterglow 
lifetime in the relativistic regime, providing a late ``energy refreshment"
to the blast wave on time scales comparable to the afterglow time scale
\cite{rm98,kupi00,dailu00,samesz00}. These examples lead to non-standard decay rates
\beq
\gamma \propto r^{-g} \propto \cases{
  r^{-(3-d)/2} & ~; $n\propto r^{-d}$;\cr
  r^{-(3-d)/(1+s)} & ~; $E(>\gamma)\propto \gamma^{1-s}~,~n\propto r^{-d}~$.\cr }
\label{eq:gammanonst}
\enq
An additional complication occurs if the outflow has a transverse ($\theta$-
dependent) gradient in its properties such as energy per solid angle or Lorentz 
factor, e.g. as some power law $\theta^{-j},~\theta^{-k}$ \cite{mrw98}. Expressions 
for the temporal decay index $\alpha (\beta,s,d,j,k,..)$ in $F_\nu\propto t^\alpha$ 
are given by \cite{mrw98,samesz00}, which now depend also on $s$, $d$, $j$, $k$, etc.
(and not just on $\beta$ as in the standard relation of equ.(\ref{eq:Fnu}).
The result is that the decay can be flatter (or steeper, depending on $s$, $d$, etc)
than the simple standard $\alpha= (3/2)\beta$,
\beq
F_\nu\propto t^\alpha\nu^\beta~~,\hbox{with}~~\alpha=\alpha (\beta,d,s,j,k,\cdots )~.
\label{eq:alphanonst}
\enq
Thus, a diversity of behaviors is not unexpected. What is more remarkable is that, 
in many cases, the simple standard relation (\ref{eq:Fnu}) is sufficient to 
describe the gross overall behavior at late times.

Strong evidence for departures from the simple standard model is provided by,
e.g., sharp rises or humps in the light curves followed by a renewed decay,
as in GRB 970508 \cite{ped98,pir98a}. Detailed time-dependent model fits
\cite{pmr98} to the X-ray, optical and radio light curves of GRB 970228 and
GRB 970508 indicate that, in order to explain the humps, a {\it non-uniform} 
injection 
is required. Other ways to get a lightcurve bump after $\sim$ days is through
microlensing \cite{garnavich00}, late injection \cite{zhang01a}, or inverse Compton
effects \cite{zhang01b,harrison01}.

\subsection{Jets and limb-brightening effects}
\label{sec:jet}
\begin{figure}
\epsfig{figure=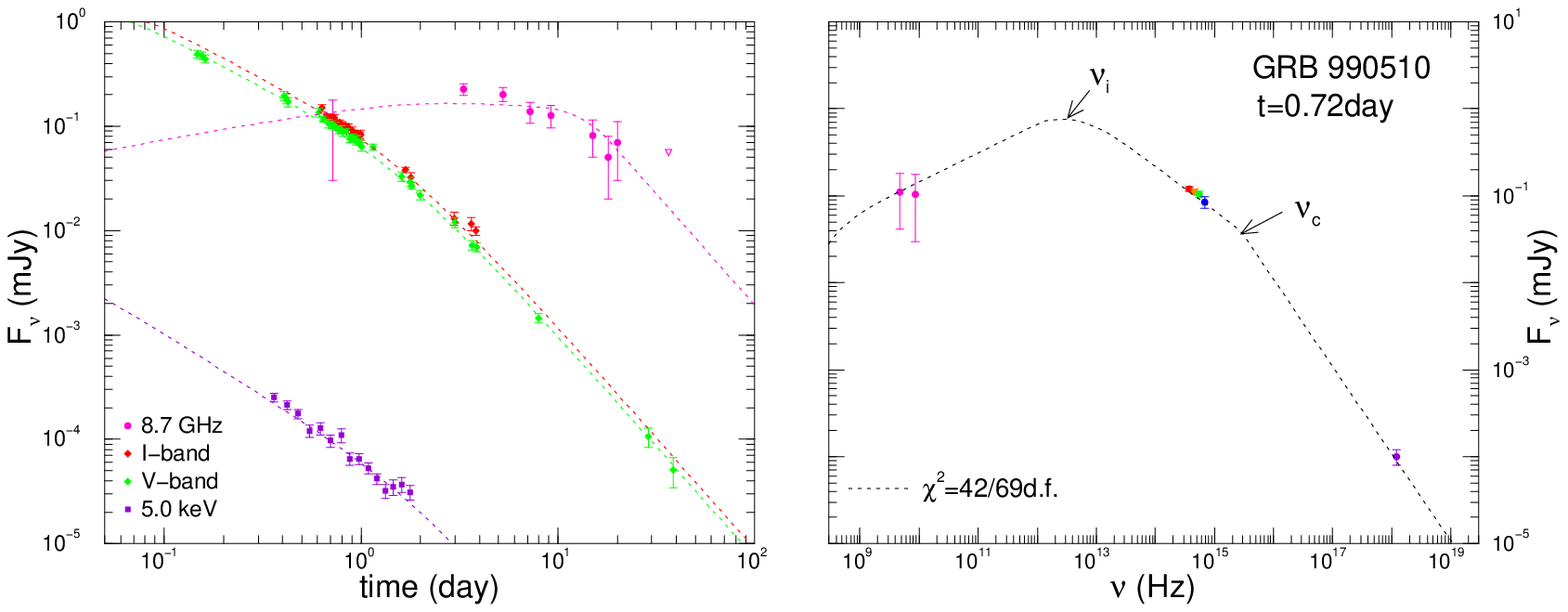,width=15cm,height=6cm} 
\caption{Model light curves at various energies (left panel) and snapshot 
spectral fit at 0.72 days (right panel) for GRB 990510, compared against the 
data \cite{pankum01a}.  The model shown has $\chi^2 = 42$ for 69 df, and 
parameters: $E_0=3.0 \times 10^{50}\;{\rm erg}$, $\theta_{jet}=2.7\deg$, 
$n_o=0.14$ cm$^{-3}$, $\eps_e=0.046$, $\eps_B=8.6 \times 10^{-4}$, $p=2.01$.
The steepening of the optical decay is due to the the effect of a jet.}
\label{fig:jet0510}
\end{figure}

The spherical assumption is valid even when considering a relativistic outflow
collimated within some jet of solid angle $\Omega_j < 4\pi$, provided the
observer line of sight is inside this angle, and $\gamma \simg \Omega_j^{-1/2}$ 
\cite{mlr93},  so the light-cone is inside the jet boundary (causally disconnected)
and the observer is unaware of what is outside the jet.
As deceleration proceeds and the Lorentz factor drops below 
this value (in $\sim$ days), a change is expected in the dynamics and 
the light curves \cite{rho97,rho99}. The first effect after $\gamma 
< \Omega_j^{-1/2}$ is that, whereas before the effective transverse emitting
area increased as $(r_\parallel/\gamma)^2 \propto t^2 \gamma^2$, thereafter it 
grows more slowly as $r_\parallel \Omega_j^{-1/2} \propto t^2 \gamma^4$, i.e. one
expects a faster decay by $\gamma^2 \propto t^{-3/4}$ \cite{mr99}, which in fact
is the magnitude of the break seen, e.g. in GRB 990123 \cite{kul99a,fru99,cas99}.
Soon after this sideways expansion of the jet would lead to an even steeper decay, 
$\propto t^{-p}$ \cite{rho97,pm98d}, possibly complicated by jet anisotropy 
\cite{daigou01}. Variable optical linear polarization can also be expected 
\cite{sari99,ghisellini99}.  An example of the lightcurve break and a snapshot 
fit is shown in Fig. \ref{fig:jet0510}. 
Numerical simulations of jet development (e.g. \citen{granot00b}) are complicated
due to the need for both high dimensionality and relativistic effects, and
comparison between such models and phenomenological fits   
\cite{fra01,pankum01a,pankum01b} still requires caution.

If the burst energy  were emitted isotropically,
the energy requirements spread over many orders of magnitude, $E_{\gamma,iso} 
\sim 10^{51}- 10^{54}$ erg \cite{kul99b}.  However, taking into account the 
evidence for jets \cite{pankum01a,pankum01b,fra01} the inferred spread in the 
total $\gamma$-ray energy is reduced to one order of magnitude, around a much 
less demanding mean value of $E_{\gamma,tot}\sim 8\times 10^{50}$ erg.  This is 
not significantly larger than the kinetic energies in core-collapse supernovae, 
although it differs from the latter by being concentrated in the gamma-ray range, 
and by being substantially more collimated than supernovae (see, however, 
\citen{whee99}).  Radiative inefficiencies and the additional energy which must 
be associated with the proton and magnetic field components increase this value, 
but it would still be well within the theoretical energetics $\siml 
10^{53.5}-10^{54}$ erg achievable in {\it either} NS-NS, NS-BH mergers \cite{mr97b} 
or in hypernova/collapsar models \cite{pac98,pop99} using MHD extraction of the
spin energy of a disrupted torus and/or a central fast spinning BH. It is worth
stressing that the presence of jets does not invalidate the usefulness of 
snapshot spectral fits, since these constrain only the {\it energy per
solid angle} \cite{mrw99}. 

An interesting property, which arises even in spherical outflows, is that the 
effective emitting region seen by the observer resembles a ring 
\cite{wax97b,goo97,pm98b,sari98,granot99a}. This effect is thought to be implicated
in giving rise to the radio diffractive scintillation pattern seen in several 
afterglows, since this requires the emitting source to be of small dimensions 
(the ring width), e.g. in GRB 970508 \cite{wkf98}. This provided 
an  important observational check, giving a direct confirmation of the 
relativistic source expansion and a direct determination of the (expected) 
source size \cite{wfk98,katz98}.

\subsection{Prompt Flashes and Reverse Shocks}
\label{sec:reverse}

A remarkable discovery was the observation \cite{ak99} of a prompt and extremely 
bright ($m_v\sim 9$) optical flash in the burst GRB 990123, 15 seconds after the 
GRB started (and while it was still going on). A prompt multi-wavelength flash,
contemporaneous with the $\gamma$-ray emission and reaching such optical magnitude 
levels is an expected consequence of the reverse component of external shocks
\cite{mr93b}. The prompt optical flash of 990123 is generally interpreted 
\cite{sp99,mr99} as the radiation from a reverse (external) shock, although a 
prompt optical flash could be expected either from an internal shock or from
the reverse external shock \cite{mr97a}. The decay rate of the optical flux
from reverse shocks is much faster (and that of internal shocks is faster still) 
than that of forward shocks, so the emission of the latter dominate after tens 
of minutes.  Such bright prompt flashes, however, appear to be rare, since they 
have not so far been detected from other bursts, either using upgraded versions 
of the original ROTSE camera \cite{kehoe01} or other similar systems 
\cite{park01,boer01}. This is further discussed by \citen{soderberg01}.

\subsection{Radiation Processes, Efficiences and Pairs }
\label{sec:pairs}

Pair-production due to $\gamma\gamma$ interactions among intra-shock photons 
satisfying $\epsilon_1 \epsilon_2 \simg m_e c^2$ can be important when the 
compactness parameter $\ell \sim n_\pm \sigma_T \Delta R_{com} >1$, where
$\Delta R_{com}$ is the comoving shock width and $n_\pm \propto L_\gamma$. This 
can affect the spectrum  of external shocks \cite{mrp94,baring00,lithwick01} 
above GeV energies. An external shock occurring beyond a preceding internal shock 
\cite{mr94} is a possible model for the EGRET GeV observations (e.g. 
\citen{hur94}) of 1-20 GeV photons in several GRBs. 
Internal shocks, occurring at smaller radii (equation [\ref{eq:rdis}]) than 
external shocks (equation [\ref{eq:rdec}]) will have larger compactness parameters, 
and pair formation can be more important \cite{rm94,pm96,pilla98}.
For close-in shocks and high luminosities, pair-breakdown could lead to 
a self-regulating moderate pair optical thickness and subrelativistic pair
temperature leading to a comptonized spectrum \cite{ghicel99}. Comptonization 
in a generic context has also been advocated by, e.g., \citen{crider97,liang99}. 

Low energy $\gamma$-ray spectral indices which appear steeper than predicted by 
a synchrotron mechanism has been reported by, e.g. \citen{preece00}. Possible 
explanations include a fireball photospheric component, photospheric bulk 
and pair-breakdown comptonization (\citen{mr00b} and references therein).
Other possibilities are synchrotron self-absorption in the X-ray \cite{granot99b}
or in the optical range upscattered to X-rays \cite{panmesz00}, low-pitch angle 
scattering \cite{medvedev00}, or time-dependent acceleration and radiation 
\cite{lloyd01b}, where the latter also point out that low-pitch angle acceleration
of electrons in a strong magnetic field may be preferred and can explain high 
energy indices steeper than predicted by an isotropic electron distribution.

A related problem is that of the radiative efficiency. For internal shocks,
this is estimated to be moderate in the bolometric sense (10-30\%), higher
values being obtained if the shells have widely differing Lorentz factors
\cite{spada00,belo00b,kobayashi01}. The total efficiency is substantially
affected by inverse Compton losses \cite{pm96,pilla98,ghis00}.  The efficiency for 
emitting in the BATSE range is typically low $\sim 2-5\%$, both when the MeV break 
is due to synchrotron \cite{kumar99,spada00,guetta01} and when it is due to
inverse Compton \cite{panmesz00}. This inefficiency is less of a concern 
when a jet is present (e.g. with typical values $\theta_{jet}\sim 3$ degrees 
and required total energies $E_0\sim 10^{50}-10^{51}$ erg, e.g.
\citen{fra01,pankum01a}).

Pair formation can also arise when $\gamma$-rays  back-scattered by the 
external medium interact with the original $\gamma$-rays \cite{derbot00}.
This may lead to a cascade and acceleration of the pairs 
\cite{thomad00,madtho00,madbr00}. For a model where $\gamma$-rays are produced
in internal shocks, analytical estimates \cite{mrr01} indicate that even for 
modest external densities a pair cloud forms ahead of the fireball ejecta,
which can accelerate to Lorentz factors $\gamma_\pm \siml 30-50$. These pairs
produce a radio signal when they are swept-up by the ejecta, and when the
pair-enriched ejecta is in turn decelerated by the external medium, its
radiative efficiency is increased. The afterglow reverse shock shares the same
energy among a larger number of leptons so that its spectrum is softened towards 
the IR (\citen{mrr01}; see also \citen{belo01}), compared of the optical/UV flash  
expected in the absence of pairs; this may contribute to the rarity of prompt 
optical detections.

Inverse Compton scattering (IC) can be an important energy loss mechanism in
external shocks \cite{mlr93} and is the likeliest mechanism for producing
GeV radiation \cite{mrp94,mr94}.  Its effects on afterglows were considered
by \cite{wax97b}, and observational manifestations in afterglows were investigated 
more carefully by \citen{pankum00} and \citen{sariesin01}. This mechanism may 
be responsible for X-ray bumps after days in some afterglow light 
curves \cite{zhang01b,harrison01}, alternative possibilities being microlensing
\cite{garnavich00} or late injection \cite{zhang01a}.

\subsection{Shock Physics}
\label{sec:shockphys}

The non-thermal spectrum in the fireball shock model is based on assuming that 
Fermi acceleration (e.g. \citen{blandford87}) accelerates electrons to highly 
relativistic energies following a power law $N(\gamma_e)\propto \gamma_e^{-p}$,
with $p\sim 2-2.5$ \cite{gallant99}. To get reasonable efficiencies, the accelerated 
electron to total energy ratio $\eps_e\siml 1$ must not be far below unity 
\cite{mr93a,kumar00}, while the magnetic to total energy ratio $\eps_b<1$ depends 
on whether the synchrotron or the IC peak represents the observed MeV break 
\cite{pm96}. The radiative efficiency and the electron power law minimum Lorentz 
factor also depends on the fraction $\zeta<1$ of swept-up electrons injected 
into the acceleration process \cite{bykov96,daigne00}.  While many 
afterglow snapshot or multi-epoch fits can be done with time-independent values 
of the shock parameters $\epsilon_b,~\epsilon_e,~p$ (e.g. \citen{wiga99}), in some
cases the fits indicate that the shock physics may be a function of the shock 
strength. For instance $p$, $\epsilon_b,~\epsilon_e$ or the electron injection 
fraction $\zeta$ may change in time \cite{pmr98,pankum01a}. While these are, 
in a sense, time-averaged shock properties,  specifically time-dependent effects 
would be expected to affect the electron energy distribution and photon spectral 
slopes, leading to time-integrated observed spectra which could differ from those 
in the simple time-averaged picture \cite{medvedev00,lloyd01a}. The back-reaction 
of protons accelerated in the same shocks (\S \ref{sec:cr}) and magnetic fields 
may also be important, as in supernova remnants (e.g. \citen{ellison00}).  
Turbulence may be important for the electron-proton energy exchange 
\cite{bykov96,schlick00}, while reactions leading to neutrons and viceversa 
\cite{rachenm98} can influence the escaping proton spectrum.

\subsection{Other Effects}
\label{sec:othereffects}

Two potentially interesting developments are the possibility of a 
relationship between the differential time lags for the arrival of the GRB pulses 
at different energies and the luminosity \cite{norris00}, and between the degree 
of variability or spikyness of the gamma-ray light curve and the luminosity 
\cite{fen00,rei00}. Attempts at modeling the spectral lags have relied on
observer-angle dependences of the Doppler boost \cite{nakamura00,salmonson01b}.
In these correlations the isotropic equivalent luminosity was used, in the absence 
of jet signatures, and they must be considered tentative for now. However, if 
confirmed, they could be invaluable for independently estimating GRB redshifts.

\section{Some Alternative Models}
\label{sec:alternativemodels}

While space limitations preclude a comprehensive review of many alternative 
models, a partial list includes precessing jets from pulsars
(\citen{blackman96}; c.f. \citen{fargion99}); jets \cite{cen97} or cannonballs 
from supernovae \cite{dar01}; magnetar bubble collapse \cite{gnedin00};
neutron star collapse to a strange star \cite{chengdai96}, or collapse to a 
black hole caused by accretion \cite{vie00} or by capture of a primordial 
black hole \cite{der99c}; supermassive black hole formation \cite{fuller98}, 
and evaporating black holes \cite{halzen91,belyanin96,cline96}.

\section{Progenitors}
\label{sec:progen}

The currently most widely held view is that GRBs arise in a very small fraction 
of stars ($\sim 10^{-6}$, or somewhat larger depending on beaming) which undergo 
a catastrophic energy release event 
toward the end of their evolution. One class of candidates involves massive stars 
whose core collapses \cite{woo93,pac98,fryerwh99}, probably in the course of merging 
with a companion, often referred to as hypernovae or collapsars. 
Another class of candidates consists of neutron star (NS) binaries or neutron 
star-black hole (BH) binaries \cite{pac86,goo86,eic89,mr97b}, which lose orbital 
angular momentum by gravitational wave radiation and undergo a merger. 
Both of these progenitor types are expected to have as an end result the formation 
of a few solar mass black hole, surrounded by a temporary debris torus whose 
accretion can provide a sudden release of gravitational energy, with similar total 
energies, sufficient to power a burst. An important point is that the overall 
energetics from these various progenitors do not differ by more than about 
one order of magnitude \cite{mrw99}. The duration of the burst in this model is
related to the fall-back time of matter to form an accretion torus around the
BH \cite{fryerwh99,pop99} or the accretion time of the torus \cite{narayan01}.  
Other possible alternatives include, e.g. the tidal disruption of compact stars by 
$10^5-10^6 \msun$ black holes \cite{bom01}, and the formation from a stellar 
collapse of a fast-rotating ultra-high magnetic field neutron star 
\cite{us94,tho94,hs99,whee00,ruderman00}.

Two large reservoirs of energy are available in such BH systems: the binding
energy of the orbiting debris \cite{woo93} and the spin energy of the black hole
\cite{mr97b}. The first can provide up to 42\% of the rest mass energy of
the disk, for a maximally rotating black hole, while the second can provide
up to 29\% of the rest mass of the black hole itself. The question is how
to extract this energy.

One energy extraction mechanisms is the $\nu\bar\nu \to e^+ e^-$ process
\cite{eic89}, which can tap the thermal energy of the torus produced
by viscous dissipation. To be efficient, the neutrinos must escape before
being advected into the hole; on the other hand, the efficiency of conversion 
into pairs (which scales with the square of the neutrino density) is low if 
the neutrino production is too gradual. Estimates suggest a  fireball of 
$\siml 10^{51}$ erg \cite{ruf97,fw98,mcfw99}, or in the collapsar case 
\cite{pop99} possibly $10^{52.3}$ ergs (c.f. higher estimates in the NS-NS 
case by \cite{salmonson01a}). If the fireball is collimated into a solid angle 
$\Omj$ then of course the apparent ``isotropized" energy would be larger by a 
factor $(4\pi/\Omj)$. Using the recent total energy estimates (corrected for jet 
collimation) $E_{\gamma,tot}\sim 10^{51}$ erg deduced from jet data by
\citen{fra01} and \citen{pankum01b}, neutrino annihilation would appear to
be a likelier possibility than it did before these analyses.
An alternative, and more efficient mechanism for tapping the energy of the
torus may be through dissipation of magnetic fields generated by the
differential rotation in the torus \cite{pac91,napapi92,mr97b,ka97}.
Even before the BH forms, a NS-NS merging system might lead to winding up of the
fields and dissipation in the last stages before the merger \cite{mr92,vie97a}.

The black hole itself, being more massive than the disk, could represent an even 
larger source of energy, especially if formed from a coalescing compact binary, 
since then it is guaranteed to be rapidly spinning. The energy extractable in 
principle through MHD coupling to the rotation of the hole by the B-Z 
\cite{bz77} mechanism could then be even larger than that contained in the 
orbiting debris \cite{mr97b,pac98}. (Less conventional and more specific 
related BH energization of jets is discussed e.g. by \citen{vanputten00}, 
\citen{li00}, \citen{ruffini01}).  Collectively, such MHD outflows have been 
referred to as Poynting jets.

The various stellar progenitors differ slightly in the mass of the BH and
somewhat more in that of the debris torus, but they can differ markedly
in the amount of rotational energy contained in the BH. Strong magnetic
fields, of order $10^{15}$ G, are needed to carry away the rotational
or gravitational energy in a time scale of tens of seconds \cite{us94,tho94},
which may be generated on such timescales by a convective dynamo mechanism,
the conditions for which are satisfied in freshly collapsed neutron stars
or neutron star tori \cite{dt92,klurud98}. If the magnetic fields do not thread 
the BH, a Poynting outflow can at most carry the gravitational binding energy
of the torus. This is
\beq
E_{t} \simeq \eps_m q 0.42 M_d c^2\siml 8\times 10^{53} \eps_m q (M_d/\msun)
                           ~\hbox{ergs}~,
\label{eq:edisk}
\enq
where $\eps_m \siml 0.3$ is the efficiency in converting gravitational into MHD 
jet energy, $q$ is in the range $[1,1/7]$ for [fast,slow] rotating BHs, and the 
mass $M_d$ of the torus or disk in a NS-NS merger is \cite{ruja98}
$\sim 10^{-1}-10^{-2}\msun$ , while in NS-BH, He-BH, WD-BH mergers or a 
binary WR collapse it may be \cite{pac98,fw98} $\sim 1\msun$.

If the magnetic fields in the torus thread the BH, the spin energy of the BH
which can be extracted e.g. through the B-Z or related mechanisms is
\cite{mr97b,mrw99}
\beq
E_{bh} \simeq \eps_m f(a)\Mbh c^2~ \siml 5\times 10^{53}\eps_m (\Mbh/\msun)~\hbox{ergs},
\enq
where $f(a)=1-([1+\sqrt{1-a^2}]/2 )^{1/2} \leq 0.29$ is the rotational efficiency 
factor, $a = Jc/G M^2=$ rotation parameter ($a=1$ for a maximally rotating BH).
The rotational factor is small unless $a$ is close to 1, so the main requirement 
is a rapidly rotating black hole, $a \simg 0.5$.  
Rapid rotation is guaranteed in a  NS-NS merger, since (especially for a soft 
equation of state) the radius is close to that of a black hole and the final
orbital spin period is close to the required maximal spin rotation period.
The central BH mass \cite{ruf97,ruja98} is $\sim 2.5 \msun$, so a NS-NS 
merger could power a jet of up to $E_{\nsns}\siml 1.3 \times 10^{54}
\eps_m$ ergs.  A maximal rotation rate may also be possible in a He-BH merger, 
depending on what fraction of the He core gets accreted along the rotation axis 
as opposed to along the equator \cite{fw98}. For a rotating He star, recent
calculations \cite{lee01} indicate that a BH rotaion parameter $a=0.7-0.9$ is
achievable. A similar end result may 
apply to the binary fast-rotating WR scenario, which probably does not
differ much in its final details from the He-BH merger. For a fast rotating
BH of $2.5-3\msun$ threaded by the magnetic field, the maximal energy
carried out by the jet is then similar or somewhat larger than in the NS-NS case.
The scenarios less likely to produce a fast rotating BH are the NS-BH merger
(where the rotation parameter could be limited to $a \leq M_{ns}/\Mbh$,
unless the BH is already fast-rotating) and the failed SNe Ib (where the
last material to fall in would have maximum angular momentum, but the
material that was initially close to the hole has less angular momentum).
Recent calculations of collapsar central BH mass/rotation rates and disk 
masses have been discussed by \citen{fryerkal01}, \citen{macfadyenwooheg01}, 
\citen{fryerwh99}, \citen{janka+99}, \citen{zhangfry01}. The magnetic interaction 
between a rotating hole and disk is further discussed in \cite{vanputost01}.
\begin{figure}
{\epsfig{figure=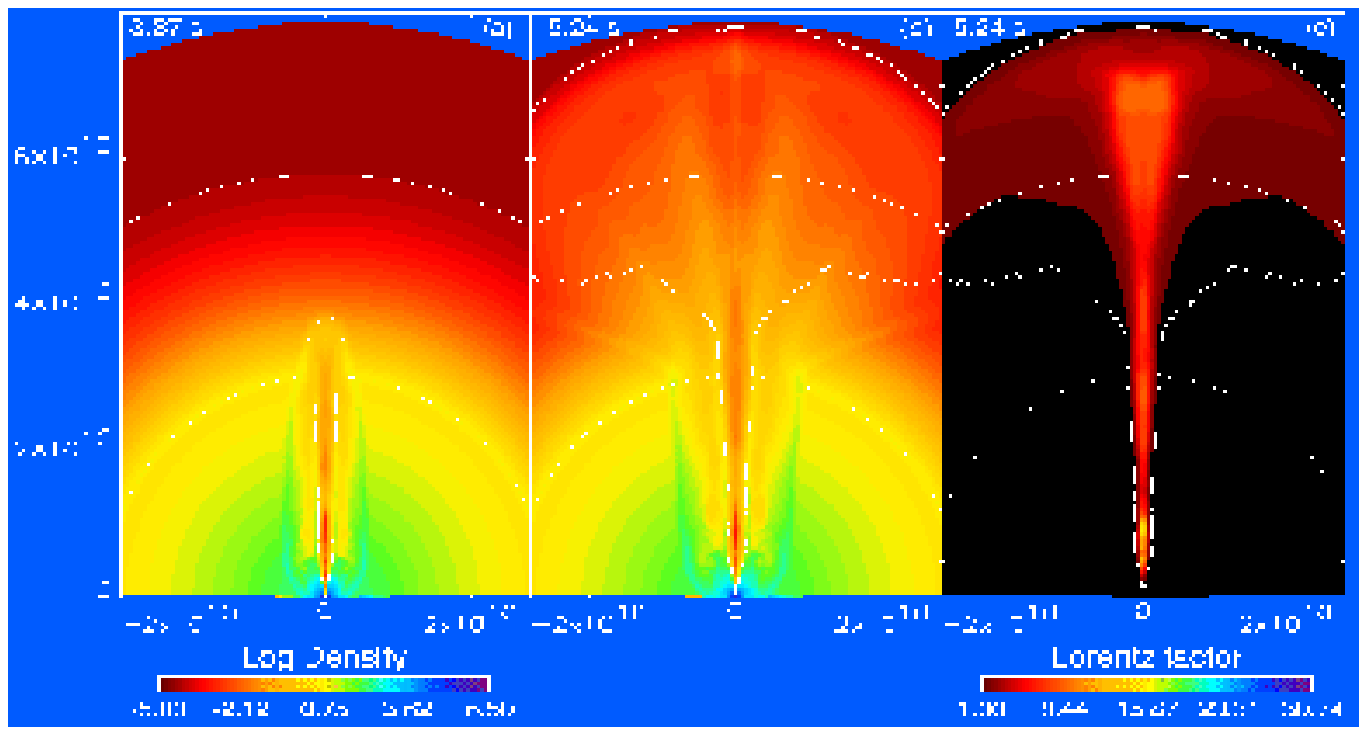,width=13cm,height=6cm} }
\caption{Jet development in a $14 \msun$ collapsar \cite{aloy00}
after substantial envelope mass loss. Contours of the logarithm of density after
3.87\,s and 5.24\,s (left two panels), and of the Lorentz factor
(right panel) after 5.24\,s. X and Y axis measure distance in
centimeters. Dashed and solid arcs mark the stellar surface and
the outer edge of the exponential atmosphere.}
\label{fig:aloyjet}
\end{figure}

The total jet energetics differ between the various BH formation scenarios at 
most by a factor 20 for Poynting jets powered by the torus binding energy,
and at most by factors  of a few for Poynting jets powered by the BH spin energy,
depending on the rotation parameter. For instance, allowing 
for a total efficiency of 50\%, a NS-NS merger whose jet is powered by the torus 
binding energy would require a beaming of the $\gamma$-rays by a factor 
$(4\pi/\Omj)\sim 100$, or beaming by a factor $\sim 10$ if the jet is powered 
by the B-Z mechanism, to produce the equivalent of an isotropic energy of 
$4\times 10^{54}$ ergs. These beaming factors are compatible with the values
derived from observations \cite{fra01} (albeit so far available for long bursts 
only).

In all cases including solar mass BHs and magnetar central objects
an $e^\pm ,\gamma$ fireball would be expected to arise from the heating and 
dissipation associated with the transient accretion event, in addition to 
MHD stresses. Even if the latter are not dominant, values in excess of $10^{15}$ 
Gauss can provide the driving stresses leading to highly relativistic $\gamma_j 
\gg 1$ expansion. The fireball would also be likely to involve some fraction of 
baryons, and uncertainties in this ``baryon pollution" \cite{pac90} remain 
difficult to dispel until 3D MHD calculations capable of addressing baryon
entrainment become available.  In spherical symmetry, general considerations 
give insights into the development of the Lorentz factor in a shock wave as it 
propagates down the density gradient of a stellar envelope \cite{sariwax00,mckee01}.
The expectation that the fireball is likely to be substantially collimated is
prevalent especially if the progenitor is a massive star, due to the constraint 
provided by an extended, fast-rotating envelope, which provides a natural fireball
escape route along the rotation axis. The development of a jet and its Lorentz 
factor in a collapsar is discussed analytically in \citen{mr01} and numerically
in, e.g. \citen{aloy00} and \citen{zhangwm01} (see Fig. \ref{fig:aloyjet})
while the case of a magnetar jet is discussed by \citen{whee00}.
In the case of \nsns or \bhns mergers a weaker degree of collimation would be 
expected, due to the lack of an extended envelope (unless magnetic or hydrodynamic 
self-collimation occurs, e.g. \citen{lev00}).

An interesting question is whether the long bursts arise from a different 
parent population as the short bursts. A current hypothesis is that while 
massive stars (e.g. via the collapsar scenario) appear implicated in long
bursts, \nsns mergers might possibly lead to short bursts \cite{katz96,pop99},
as also discussed in the next \S. (c.f. \citen{vanputost01} for an alternative 
view in which both long and short bursts originate in collapsars).

\section{Cosmological Setting, Galactic Hosts and Environment}
\label{sec:cosm}

For the long GRB afterglows localized so far, a host galaxy has been found in most 
cases ($\simg 80\%$ out of over 30 optically identified, \citen{bloom01}). The 
GRB hosts are typically low mass, sub-$L_\ast$ galaxies, with the blue colors and 
atomic lines indicative of active star formation (\citen{fru00,bloom01,fra01}; see 
also \citen{scha00}). Many of them are obscured, far-infrared luminous galaxies, 
some of which appear tidally disturbed \cite{chary01}.  The redshifts of the hosts, 
with one exception, are in the range $0.43 \siml z \siml 4.5$, 
i.e., comparable to that of the most distant objects detected in the Universe.
The observed number of bursts per unit photon flux can be fitted by cosmological 
distribution models, with a somewhat better fit if one assumes that the burst rate 
scales proportionally to the observed star-formation rate as a function of 
redshift \cite{wi98,tot99,bn00,bd00,stern01}. The spread in the inferred 
isotropic-equivalent luminosities extends over three orders of magnitude, i.e.
far from standard candles for the purposes of testing cosmological models 
\cite{maomo98}.  However, this spread in luminosities is considerably reduced 
to less than one order of magnitude \cite{pankum01a,fra01,pankum01b,piran01} if 
allowance is made for jet-like collimation. The sample of bursts for which this 
is possible is still too small ($\siml 10$) to do cosmology with them.

The bursts for which the intrinsic brightness is known from their measured
redshifts would, in principle, be detectable out to much larger redshifts
$z \siml 15-20$ with present detectors \cite{lr00}. Within the first minutes 
to hours after the burst, the afterglow optical light is expected to be in the
range $m_v \sim 10-15$, far brighter than quasars, albeit for a short time. 
Thus, promptly localized GRB could serve as beacons which, shining through the 
pregalactic gas, provide information about much earlier epochs in the history of 
the Universe. The presence of iron or other x-ray lines provides an additional tool 
for measuring GRB distances, which may be valuable for investigating the small but 
puzzling fraction of bursts which have been detected only in X-rays but not 
optically, perhaps due to a high dust content in the host galaxy.

Accurate localizations and host galaxies have, so far, been restricted to the class
of ``long" bursts ($\gamma$-ray durations $t_b \sim 10-10^3$ s), because BeppoSAX 
is mostly sensitive to bursts longer than about 5-10 s. (One exception is a recent 
short burst localization, which led to optical upper limits $R > 22.3$ and 
$I > 21.2$ about 20 hours after the trigger \citen{gorosabel01}. For the long 
bursts, the fading x-ray and optical afterglow emission is predominantly localized 
within the optical image of the host galaxy.  In most cases it is offset from the 
center, but in a few cases (out of a total of about twenty) it is near the center 
of the galaxy \cite{bloom01}. This is in disagreement with current simple 
calculations of NS-NS mergers which suggest (\citen{bsp99}; also \citen{napapi92}) 
that high spatial velocities would take these binaries, 
in more than half of the cases, outside of the confines of the host galaxy before 
they merge and produce a burst. These calculations, however, are uncertain, since 
they are sensitive to a number of poorly known parameters (e.g distribution of 
initial separations, etc).  On the other hand, theoretical estimates \cite{fryerwh99}
suggest that NS-NS and NS-BH mergers will lead to shorter bursts ($\siml 5 s$), 
beyond the capabilities of Beppo-SAX but expected to be detectable with the 
recently launched HETE-2 spacecraft \cite{hete00}. More effectively, short as 
well as long bursts should be detected at the rate of 200-300 per year with the 
{\it Swift} multi-wavelength GRB afterglow mission \cite{swift00} currently 
under construction and scheduled for launch in 2003.  Swift will be equipped 
with $\gamma$-ray, x-ray and optical detectors for on-board follow-up, and will
capable to slew within 30-70 seconds its arc-second resolution X-ray camera onto 
GRBs acquired with their large field-of-view gamma-ray monitor, relaying to the 
ground the burst coordinates within less than a minute from the burst trigger.
This will permit much more detailed studies of the burst environment, the host 
galaxy, and the intergalactic medium. 

Hydrogen Lyman $\alpha$ absorption from intervening newly formed galaxies would 
be detectable as the GRB optical/UV continuum light shines through them 
\cite{lb01,lr00}.  While the starlight currently detected is thought to come 
mostly from later, already metal-enriched generations of star formation, GRB 
arising from the earliest generation of stars may be detectable; and if this 
occurs before galaxies have  gravitationally assembled, it would provide a 
glimpse into the pregalactic phase of the Universe.
At a given observed wavelength and a given observed time delay, the observed 
brightness of a burst afterglow decreases more slowly at higher redshifts,
since the afterglow is observed at an earlier source time and at a 
higher frequency where it is brighter \cite{ciardi00,lr00}. The high redshift
afterglows shining through their host or intervening galaxies would be expected 
to provide valuable information in the near IR, while in the far IR and sub-mm 
they would provide invaluable information about the dust content in high
redshift environments \cite{blain01}. Dust affects the colors of the light
curves and contains information about the metallicity as a function of redshift
\cite{reichart01}.  Bursts which are highly dust-obscured in
the optical would generally be detectable in $\gamma$-rays and X-rays, and
quantitative information about the dust content may be obtained through the 
detection of a hump accompanied by a spectral softening in the keV X-ray light 
curve \cite{meszgruz00}, caused by small-angle forward scattering on the dust 
grains, accompanied by a late brightening in the near-IR. 

Most of the host galaxies of the long bursts detected so far show signs of 
active star formation, implying the presence of young, massive stars forming 
out of dense gaseous clouds.  The diffuse gas around a GRB is expected to produce 
time-variable O/UV atomic absorption lines in the first minutes to hours after 
a burst \cite{pl98}.  There is also independent evidence from the observation 
of 0.5-2 keV absorption in the x-ray afterglow spectra, attributed to metals in 
a high column density of gas in front of the burst \cite{gawi01}. This appears
to be higher than expected from optical extiction measures, which may be due
to dust destruction by UV photons \cite{waxdrai00,esinbla00,fru01}.

It is interesting that, at least in a few bursts so far, there appears to 
be evidence for an approximately coincident supernova explosion. 
There is good spatial-temporal coincidence for one burst, GRB 980425, 
associated with the unusually bright SN Ib/Ic 1998bw 
\cite{gal98_SN,bloom98_SN,jvp99}. At a measured redshift  of 0.0085 the 
association would imply an abnormally faint GRB luminosity ($\sim 10^{47}$ ergs), 
although it can be argued that the jet appears fainter due to being seen by 
chance almost close to its edge (e.g. \citen{whee99}). 
For SN 1998bw, a mildly relativistic and quasi-spherical shock break-out is also
a good model \cite{wl99,mckee01}.
In at least three other localized long 
GRB, there is circumstantial evidence for a supernova remnant in the form of a bump 
and reddening in the GRB afterglow optical light curve after several weeks
\cite{lazzati01b,galama00_SN,reichart99_SN,bloom99_SN}. Alternative explanations
based on dust sublimation and scattering have been proposed by \citen{esinbla00}
and \citen{waxdrai00}. The hypothesis of a generic association of GRB and 
supernovae (``hypernovae") has been discussed by \citen{pac98} (see also 
\citen{woo93}) and by \citen{whee00}, while multiple sub-jets are discussed 
by \cite{nakamura00}.

X-ray atomic edges and resonance absorption lines are expected to be detectable 
from the gas in the immediate environment of the GRB, and in particular from the 
remnants of a massive progenitor stellar system \cite{mr98b,weth00,botfry01}. 
Observations with the {\it Chandra} ACIS X-ray spectrographic camera has provided 
evidence, at a moderate $\simg 4\sigma$ confidence level, for iron K-$\alpha$ line 
and edge features in at least one burst (GRB 991216, \citen{piro00}), and there are 
at least four other detections at the $\sim 3\sigma$ level with Beppo-SAX and ASCA 
(e.g. \citen{amati00}, \citen{yoshida99}, \citen{jvp00}). The observed frequency of 
the iron lines appear displaced from the laboratory frequency by the right amount 
expected from the measured optical redshift, when available, indicating that the 
material producing the lines is expanding at $v/c \siml 0.1$ \cite{piro00}. 
The presence of iron line features would again strongly suggest a massive stellar 
progenitor, but the details remain model dependent. 

\begin{figure}[ht]
\centering
\epsfig{file=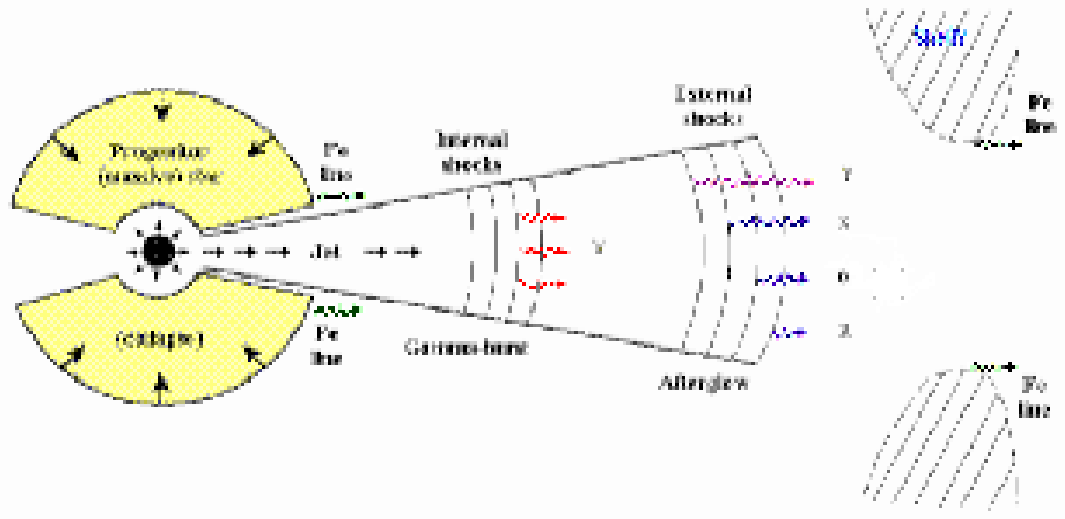,width=4.5in,height=2.7in} 
\vspace*{-1cm}
\caption{
Schematic GRB from internal shocks and afterglow from external shock, arising 
from a relativistic jet emerging from a massive  progenitor collapse (similar 
jets could arise from other progenitors). Internal shocks produce $\gamma$-rays 
and PeV neutrinos, external shocks produce $\gamma$-rays, X-rays, optical, 
radio and EeV neutrinos. Fe X-ray lines may arise from X-ray illumination of 
a pre-ejected supernova remnant \cite{piro00} or from continued X-ray 
irradiation of the outer stellar envelope \cite{rm00} (c.f. also 
Fig. \ref{fig:bubble}). }
\label{fig:collaps}
\end{figure}

One possible interpretation of the iron emission lines ascribes the approximate 
one  day delay between the burst and the Fe line peak to light-travel time 
effects, a specific example postulating an Fe-enriched supernova remnant situated 
outside the burst region, which is illuminated by X-rays from the afterglow leading 
to Fe recombination line emission (Fig. \ref{fig:collaps}).  
This would require about $10^{-1}-1~\msun$ of Fe in the shell from, e.g., a
supernova (SN) explosion by the progenitor occurring weeks before the burst, 
which might be due from the accretion-induced collapse of the NS remnant left 
behind by the SN \cite{piro00,vie00}. A similar interpretation is made 
\cite{lazzati01c} in the one reported Beppo-SAX case which appears as prompt 
($\siml 40$ s) Fe absorption feature \cite{amati00}.  A delay of weeks is required 
to allow SNR shell to travel out to a light-day distance and for the Ni in the 
explosion to decay to Fe. If the lines are ascribed to Ni or Co \cite{lazzati01a} 
the shell velocity must match the difference to the Fe line energies. In either 
case some fine-tuning appears necessary. 

\begin{figure}[ht]
\centering
\vspace*{-1cm}
\epsfig{file=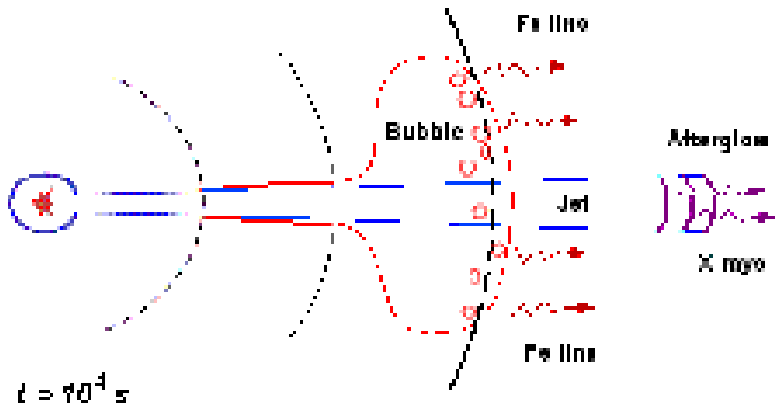,width=4.2in,height=2.5in}
\vspace*{-1cm}
\caption{
Schematic GRB afterglow from a jet emerging from a massive progenitor, 
followed hours later by emergence of a bubble of waste heat producing 
additional non-thermal X-ray and reprocessed Fe K$\alpha$ recombination 
from dense filaments in the bubble and envelope \cite{mr01}.}
\label{fig:bubble}
\end{figure}

A less demanding Fe line model is possible if the GRB, after its usual initial 
outburst, continues to eject a progressively weaker jet for a few days, at a 
rate which does not violate the observed light-curve \cite{rm00}.  This jet may
be fed, e.g. through continued fall-back at low on the BH, or through spin-down 
if the central object is a magnetar.  A decaying jet with a luminosity $L\sim 
10^{47}$ erg/s at one day impinging on the outer layers near $\sim 10^{13}$ cm
of the progenitor envelope leads to Fe recombination line emission at the observed 
rate, requiring only solar abundances or a total of $\sim 10^{-5}\msun$ of Fe
(Fig. \ref{fig:collaps}). 
However, the most plausible model may be one based upon the the after-effects of 
the cocoon of waste heat pumped into the lower envelope as the relativistic jet 
makes its way through the progenitor envelope \cite{mr01}. This bubble 
(Figure \ref{fig:bubble})
of waste heat, after the jet has emerged and produced the burst, rises slowly by 
buoyancy and emerges through the outer envelope on timescales of a day after 
the burst. Its structure is likely to be highly inhomogeneous, resulting in
non-thermal X-rays produced by synchrotron in the low density medium between 
much denser photo-ionized filaments which can produce the observed Fe line 
luminosity through recombination, requiring a modest $\sim 10^{-5}\msun$ of Fe
which can be easily supplied from the core of the star as the jet develops.
In this type of nearby ($\siml 10^{13}$ cm) line production models, the Fe line
energies could be more naturally mimicked by down-scattering of Ni or Co lines
\cite{mclaughlin01}.

The simple picture of an origin in star-forming regions, at least for the long
($t_b \simg 5$ s) bursts, is complicated by the fact that the observed optical 
absorption is less than expected for the corresponding x-ray absorption.  Also, 
standard afterglow model fits indicate an ambient gas density generally lower 
than that expected in star-forming clouds \cite{gawi01,pankum01b}. These 
contradictions may possibly be reconcilable, e.g. through dust sublimation 
by x-ray/UV radiation \cite{waxdrai00,esinbla00,fru01} or the blowing out of 
a cavity by a progenitor wind \cite{wijers00}.

While it is unclear whether there is one or more classes of GRB progenitors,
e.g. corresponding to short and long bursts, there is a general consensus that they 
would all lead to the generic fireball shock scenario. Much of the current effort 
is dedicated to understanding the different progenitor scenarios, and trying to 
determine how the progenitor and the burst environment can affect the observable 
burst and afterglow characteristics.

\section{Cosmic Rays, Neutrinos, GeV-TeV Photons and Gravity Waves } 

\label{sec:cr}

There are several other, as yet unconfirmed, but potentially interesting observing
channels for GRBs, relating to the baryonic  component of the outflow, the shock 
physics and the progenitor collapse dynamics.

Among these, cosmic rays are perhaps most directly implicated in the fireball 
shock mechanism, thought to accelerate the electrons responsible for the 
non-thermal $\gamma$-rays in GRB. The same shocks should also accelerate protons, 
based on experience from interplanetary shocks. 
Both the  internal and the external reverse shocks are mildly relativistic, and 
are expected to lead to relativistic proton energy spectra of the form 
$d N_p/d \eps_p \propto \gamma_p^{-2}$ (\citen{blandford87}; see also
\citen{kirk00} and \citen{lloyd01b}). The maximum proton energies achievable in GRB 
shocks are $E_p\sim 10^{20}$ eV \cite{wax95,vie95,dm01}, comparable to the highest 
energies measured with large cosmic ray ground arrays (e.g. \citen{ag99}).
The condition for this is that the acceleration time up to that energy is shorter 
than the radiation or adiabatic loss time as well as the escape time from the 
acceleration region. The resulting constraints on the magnetic field and the bulk 
Lorentz factor \cite{wax95} are close to those required to obtain efficient 
gamma-ray emission at $\sim 1$ MeV. If the accelerated electrons which produce the 
$\gamma$-rays and the protons carry a similar fraction of the total energy
(a conservative assumption, based on interplanetary collisionless shock 
acceleration measurements), the GRB cosmic ray energy production rate at 
$10^{20}$ eV throughout the universe is of order $10^{44}$ erg/Mpc$^3$/yr, 
comparable to the observationally required rate from $\gamma$-ray observations 
and from the observed diffuse cosmic ray flux (\citen{wax95}; c.f. \citen{ste00a}). 
These numbers depend on uncertainties in the burst total energy and beaming 
fraction, as well as on the poorly constrained burst rate evolution with redshift. 
The highest energy 
protons would need to have arrived from within less than about 50-100 Mpc, to
avoid interaction with the microwave background, and reasonable intergalactic 
magnetic field strengths can ensure time dispersions in excess of a few hundred 
years, needed to achieve compatibility with the estimated burst rate of $\sim 
10^{-6}$ /galaxy/year, as well as with arrival from clustered sources \cite{bw00}. 
The unknown strength and correlation length of the field could lead to 
anisotropies constraining both GRB models and competing AGN or other discrete 
source origin models, an issue which will be addressed by future large area 
ground cosmic ray arrays such as, e.g., Auger and HiRes. 

Any stellar origin mechanism (whether collapsar, neutron star merger, etc) would
lead to a very large ($\sim \msun c^2$) luminosity in thermal neutrinos and 
antineutrinos with energies  $\sim$ few MeV, as in core-collapse supernova. 
However, at MeV energies the neutrino detection cross section is $\sim 10^{-44}$ 
cm$^2$, and as shown by the low count rates in the supernova SN 1987a detection from
$\sim 50$ Kpc, even larger detectors at these energies (super-Kamiokande, Sudbury, 
etc) would be insensitive to sources such as GRB with typical distances $\simg$ 
100 Mpc. 

A mechanism leading to higher (GeV) energy neutrinos in GRB is inelastic nuclear 
collisions. Proton-proton collisions at internal shock radii $\sim 10^{14}$ cm 
could lead to $\sim$ GeV neutrinos in the observer frame through charged pion 
decay \cite{px94}, with low efficiency due to the low densities at these large
radii and small relative velocities between protons. However, proton-neutron 
inelastic collisions are expected, even in the absence of shocks, at much lower 
radii, due to the decoupling of neutrons and protons in the fireball or jet 
\cite{der99a}. Provided the fireball has a substantial neutron/proton ratio, as 
expected in most GRB progenitors, the collisions become inelastic and their rate 
peaks at when the nuclear scattering time becomes comparable to the expansion time.  
This occurs when the $n$ and $p$ fluids decouple, their relative drift velocity 
becoming comparable to $c$, which is easier due to the lack of charge of the
neutrons. Inelastic $n,p$ collisions  then lead to charged pions and GeV 
muon and electron neutrinos \cite{bm00}. The early decoupling and saturation of 
the $n$ also leads to a somewhat higher final $p$ Lorentz factor 
\cite{der99a,bm00,fuller00,pruet01}, implying a possible relation between the 
$n/p$ ratio and the observable fireball dynamics, relevant for the progenitor 
question and burst timescales.  Inelastic $p,n$ collisions leading to neutrinos 
can also occur in fireball outflows with transverse inhomogeneities in the bulk 
Lorentz factor, where the $n$ can drift sideways into regions of different bulk 
velocity flow, or in situations where internal shocks involving $n$ and $p$ occur 
close to the saturation radius or below the photon photosphere \cite{mr00a}.  
The typical $n,p$ neutrino energies are in the 5-10 GeV range, which could be 
detectable in coincidence with observed GRBs for a sufficiently close  photo-tube 
spacing in future km$^3$ detectors such as ICECUBE \cite{halzen00}.
\begin{figure}[ht]
\centering
\epsfig{file=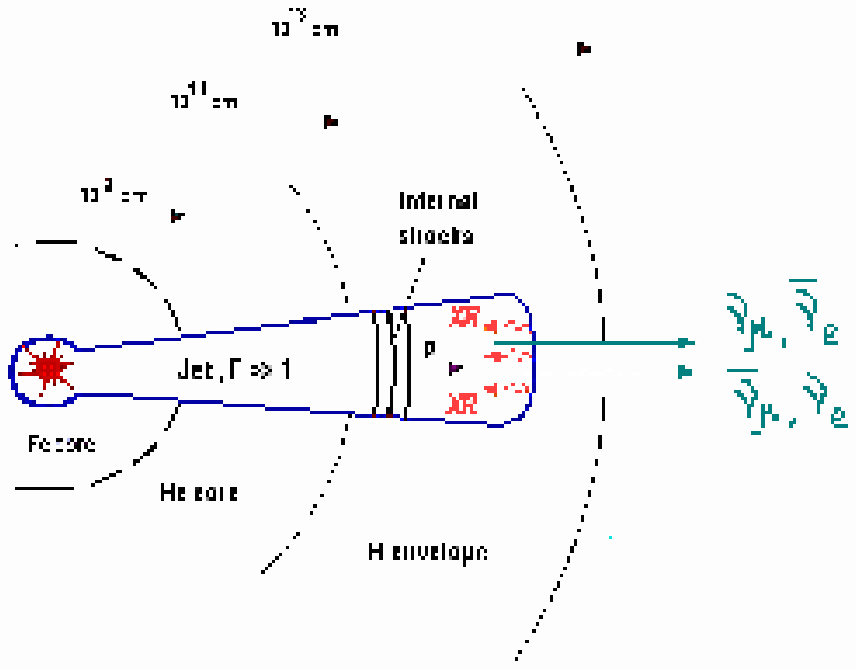,width=4.0in,height=3.0in}
\caption{Sketch of TeV neutrino production by photomeson interactions  in
internal shocks before a relativistic jet has broken through the progenitor 
envelope \cite{mw01}. Neutrinos would be expected whether the jet is choked
off ($\gamma$-dark) or emerges to make a GRB. }
\label{fig:jetnu}
\end{figure}

In addition, neutrinos with energies $\simg$ PeV  can be produced in $p,\gamma$ 
photo-pion interactions involving highly relativistic protons accelerated in the
fireball internal or external shocks. A high collision rate is ensured here
by the large density of photons in the fireball shocks. The most obvious case 
is the interaction between MeV photons produced by radiation from electrons
accelerated in internal shocks (see Fig. \ref{fig:collaps}), 
and relativistic protons accelerated by the 
same shocks \cite{wb97}, leading to charged pions, muons and neutrinos. This 
$p,\gamma$ reaction peaks at the energy threshold for the photo-meson $\Delta$ 
resonance in the fluid frame moving with $\gamma$, or
\beq
\eps_p\eps_\gamma \simg 0.2 \hbox{GeV}^2 \gamma^2~.
\label{eq:pgamma}
\enq
For observed 1 MeV photons this implies 
$\simg 10^{16}$ eV protons, and neutrinos with $\sim 5\%$ of that energy, 
$\eps_\nu\simg 10^{14}$ eV in the observer frame. Above this threshold, the 
fraction of the proton energy lost to pions is $\sim 20\%$ for typical fireball 
parameters, and the typical spectrum of neutrino energy per decade is flat, 
$\eps_\nu^2 \Phi_\nu \sim$ constant. Synchrotron and adiabatic losses limit the 
muon lifetimes \cite{rachenm98}, leading to a suppression of the neutrino flux 
above $\eps_\nu \sim 10^{16}$ eV. In external shocks (Fig. \ref{fig:collaps}),
another copious source of 
targets are the O/UV photons in the afterglow reverse shock (e.g. as deduced from
the GRB 990123 prompt flash of \citen{ak99}). In this case the resonance condition 
implies higher energy protons, leading to neutrinos of $10^{17}-10^{19}$ eV 
\cite{wb99a,vie98}. These neutrino fluxes are expected to be detectable above 
the atmospheric neutrino background with the planned cubic kilometer ICECUBE 
detector \cite{halzen00,alvarez00}. Useful limits to their total contribution to 
the diffuse ultra-high energy neutrino flux can be derived from observed cosmic 
ray and diffuse gamma-ray fluxes \cite{wb99b,bw01,mann01}.
While the $p,\gamma$ interactions leading to $\simg 100$ TeV energy neutrinos 
provide a direct probe of the internal shock acceleration process, as well as 
of the MeV photon density associated with them, the $\simg 10$ PeV neutrinos 
would probe the reverse external shocks, as well as the photon
energies and energy density there. 

The most intense neutrino signals, however, may be due to $p,\gamma$ interactions
occurring {\it inside} collapsars while the jet is still burrowing its way out of 
the star \cite{mw01}, before it has had a chance to break through (or not) the 
stellar envelope to produce a GRB outside of it.  While still inside the star, the 
buried jet produces thermal X-rays at $\sim 1$ keV which interact with $\simg 10^5$ 
GeV protons which could be accelerated in internal shocks occurring well inside
the jet/stellar envelope terminal shock, producing $\sim$ few TeV neutrinos for
tens of seconds, which penetrate the envelope (Figure \ref{fig:jetnu}).
This energy is close to the maximum sensitivity for detection, and 
the number of neutrinos is also larger for the same total energy output. The rare 
bright, nearby or high $\gamma$ collapsars could occur at the rate of $\sim 
10$/year, including both $\gamma$-ray bright GRBs (where the jet broke through the 
envelope) and $\gamma$-ray dark events where the jet is choked (failed to break 
through), and both such $\gamma$-bright and dark events could have a TeV neutrino 
fluence of $\sim 10$/neutrinos/burst, detectable by ICECUBE in individual bursts. 

GeV to TeV photon production is another consequence of the photo-pion and 
inelastic collisions responsible for the ultra-high energy neutrinos 
\cite{wb97,bd98,der99a,bm00}. This is in addition to the GeV emission from 
electron inverse Compton in internal \cite{pm96} and external shocks 
\citen{mrp94,der01} and afterglows \cite{zhang01b}. In these models, due to 
the high photon densities implied by GRB models, $\gamma\gamma$ absorption within 
the GRB emission region must be taken into account (see also 
\citen{baring00,lithwick01}).  A tentative 
$\simg 0.1$ TeV detection of an individual GRB has been reported with the water 
Cherenkov detector Milagrito \cite{atk00},  and better sensitivity is expected
from its larger version MILAGRO as well as from atmospheric Cherenkov telescopes 
under construction such as VERITAS, HESS, MAGIC and CANGAROO-III \cite{weekes00}.
GRB detections in the TeV range are expected only for rare nearby events, since
at this energy the mean free path against $\gamma\gamma$ absorption on the diffuse 
IR photon background is $\sim$ few hundred Mpc \cite{coppi97,ste00b}.
The mean free path is much larger at GeV energies, and based on the handful of GRB 
reported in this range with EGRET \cite{schneid95}, several hundred should be 
detectable with large area space-based detectors such as GLAST 
\cite{glast00,zhang01b}, in coincidence with the neutrino pulses and the usual 
MeV $\gamma$-ray event. Their detection would provide important constraints on 
the emission mechanism and the progenitors of GRBs.

GRB are also expected to be sources of gravitational waves. A time-integrated 
luminosity of the order of a solar rest mass ($\sim 10^{54}$ erg) is predicted 
from merging NS-NS and NS-BH models \cite{napapi92,kp93,ruja98,nakamura99}, while the 
luminosity from collapsar models is more model-dependent, but expected to be lower 
(\citen{fryerwh01,muellergw01}; c.f. \citen{vanputten01}). 
The rates of gravitational wave events \cite{finn00} detectable by the Laser 
Interferometric Gravitational Wave Observatory (LIGO, currently under construction) 
from compact binary mergers, in coincidence with GRBs, has been estimated at a 
few/year for the initial LIGO, and up to 10-15/year after the upgrades planned 
2-4 years after first operations.  The observation of such gravitational waves 
is greatly facilitated by  coincident  detections in other channels, either
electromagnetic or neutrinos. Detection of gravity wave pulses fitting the templates
for compact binary mergers (or collapsars), in coincidence with  positive GRB 
localizations, would have a great impact on our understanding of GRB progenitor 
systems. 

In conclusion, major advances have been made in the understanding of GRB since 
their discovery almost 30 years ago. However many questions remain, while new
ones arise in the wake of the increasingly sophisticated and extensive observations.
These questions will be addressed with new space missions and ground experiments 
dedicated to GRB studies which will come on-line in the near future. Based on past
experience the chances are high that these will bring not only answers but 
also new surprises and challenges. 

{\it Acknowledgments:}{It is a pleasure to thank M.J. Rees, as well as
C. Dermer, V. Kocharovsky, R. Narayan, R. Sari, E. Waxman, R. Wijers, B. Zhang for 
collaborations and/or comments and NASA NAG5-9192, NSF AST 0098416 for support.}

\eject





\end{document}